# On Generalizations of the Minimal Complementary Energy Variational Principle in Linear Elastostatics

Jiashi Yang (jyang1@unl.edu)
Department of Mechanical and Materials Engineering
University of Nebraska-Lincoln, Lincoln, NE 68588-0526, USA

**Abstract**
It is shown that when the well-known minimal complementary energy variational principle in linear elastostatics is written in a different form with the strain tensor as an independent variable and the constitutive relation as one of the constraints, the removal of the constraints by Lagrange multipliers leads to a three-field variational principle with the displacement vector, stress field and strain field as independent variables. This three-field variational principle is without constrains and its variational functional is different from those of the existing three-field variational principles. The generalization is not unique. The procedure is mathematical and may be used in other branches of physics.

## 1. Introduction

The principle of minimal complementary energy in linear elastostatics is well known [1,2]. Its independent variable is the stress field only, with the equilibrium equation and the traction boundary condition as constraints. If the constraints are removed using Lagrange multipliers, the principle of minimal complementary energy leads to the well-known Hellinger-Reissner variational principle [3,4] whose independent variables are the stress and displacement fields without constraints.

In comparison, it also well-known that from the one-field (displacement) principle of minimal potential energy [1], with the removal of the constraints by Lagrange multipliers, a three-field variational principle (the Hu-Washizu principle [1,2,5]) can be obtained with the displacement, stress and strain fields as independent variables and it does not have constraints.

A natural question is how to obtain from the principle of minimal complementary energy a three-field variational principle. Some researchers hold an opinion that the stress-strain relation is also a constraint of the principle of minimal complementary energy. However, this constraint cannot be removed by conventional Lagrange multipliers and some higher-order Lagrange multipliers maybe used to resolve this problem [2]. [2] has motivated some further studies [6].

In this paper we show that when the principle of the minimal complementary energy is properly written in a different form with the strain tensor as an independent variable and the constitutive relation as one of the constraints, all constraints can be removed by conventional Lagrange multipliers which results in a three-field variational principle with the displacement vector, stress field and strain field as independent variables. This three-field variational principle is without constrains and its variational functional is different from those of the existing three-field variational principles.

## 2. Governing Equations and Boundary Conditions

We follow the notation of [2]. Let the displacement vector, stress tensor and strain tensor be **u**, **σ** and **e**, respectively. The governing equations of linear elastostatics are the equilibrium equation, the strain-displacement relation and the constitutive relation. The equilibrium equation is

$$\sigma_{ji,j} + \bar{F}_i = 0, \qquad (1)$$

where $\bar{F}_i$ is the body force per unit volume and $\sigma_{ji}=\sigma_{ij}$. The strain-displacement relation is

$$e_{ij} = \frac{1}{2}(u_{i,j}+u_{j,i}). \tag{2}$$

The constitutive relation may be written as either

$$\sigma_{ij} = a_{ijkl}e_{kl}, \tag{3}$$

or

$$e_{ij} = b_{ijkl}\sigma_{kl}, \tag{4}$$

where $a_{ijkl}$ and $b_{ijkl}$ are the elastic stiffness and compliance, respectively, and one is the inverse of the other. We also have

$$\frac{1}{2}\sigma_{ij}e_{ij} = \frac{1}{2}a_{ijkl}e_{ij}e_{kl} = \frac{1}{2}b_{ijkl}\sigma_{ij}\sigma_{kl}, \tag{5}$$

$$\sigma_{ij}e_{ij} = \frac{1}{2}\sigma_{ij}e_{ij}+\frac{1}{2}\sigma_{ij}e_{ij} = \frac{1}{2}a_{ijkl}e_{ij}e_{kl}+\frac{1}{2}b_{ijkl}\sigma_{ij}\sigma_{kl}. \tag{6}$$

Consider an elastic body occupying a spatial region $V$ whose boundary surface is $S$ with an outward unit normal **n**. Let $S$ be partitioned into two parts of $S_u$ and $S_\sigma$ such that

$$S_u \cup S_\sigma = S, \quad S_u \cap S_\sigma = \varnothing. \tag{7}$$

Typical boundary conditions are

$$u_i = \bar{u}_i \quad \text{on} \quad S_u, \tag{8}$$

$$n_j\sigma_{ji} = \bar{p}_i \quad \text{on} \quad S_\sigma, \tag{9}$$

where $\bar{u}_i$ and $\bar{p}_i$ are prescribed boundary displacement and traction.

## 3. The Principle of Minimal Complementary Energy and the Hellinger-Reissner Principle

The variational functional of the principle of minimal complementary energy is [2]

$$\Pi_C(\boldsymbol{\sigma}) = \int_V \frac{1}{2}b_{ijkl}\sigma_{ij}\sigma_{kl}dV - \int_{S_u}\bar{u}_i n_j\sigma_{ji}dS, \tag{10}$$

with the following constraints:

$$\sigma_{ji,j}+\bar{F}_i = 0 \quad \text{in} \quad V, \tag{11}$$

$$n_j\sigma_{ji} = \bar{p}_i \quad \text{on} \quad S_\sigma. \tag{12}$$

If Lagrange multipliers are used to remove the constraints in (11) and (12), it results in the Hellinger-Reissner Principle whose functional is given by [2]

$$\Pi_{HR}(\boldsymbol{\sigma},\mathbf{u}) = \int_V [-\frac{1}{2}b_{ijkl}\sigma_{ij}\sigma_{kl} - u_i(\sigma_{ji,j}+\bar{F}_i)]dV$$
$$+\int_{S_u}\bar{u}_i n_j\sigma_{ji}dS + \int_{S_\sigma}u_i(n_j\sigma_{ji}-\bar{p}_i)dS. \tag{13}$$

(13) has two independent fields and does not have constraints. The stationary conditions of (13) are

$$\sigma_{ji,j}+\bar{F}_i = 0 \quad \text{in} \quad V, \tag{14}$$

$$\frac{1}{2}(u_{i,j}+u_{j,i}) = b_{ijkl}\sigma_{kl} \quad \text{in} \quad V, \tag{15}$$

$$u_i = \bar{u}_i \quad \text{on} \quad S_u, \tag{16}$$

$$n_j\sigma_{ji} = \bar{p}_i \quad \text{on} \quad S_\sigma. \tag{17}$$

The above are well-documented in the literature. (14)-(17) represent a well-defined boundary-value problem in linear elastostatics. (15) is equivalent to the elimination of **e** using (2) and (4). Hence **e** is absent in the formulation of the theory of elasticity in terms of (13) or (14)-(17).

### 4. Generalizations of the Principle of Minimal Complementary Energy

The Hellinger-Reissner principle may be viewed as a generalization of the principle of minimal complementary energy. However, different from the generalization of the principle of minimal potential energy to the Hu-Washizu principle which is a three-field principle with all of **u**, **σ** and **e** as independent variables, the Hellinger-Reissner principle is a two-field principle in terms of **u** and **σ** only without **e**. [2] believes that the stress-strain relation (4) is a constraint of the principle of minimal complementary energy but this constraint cannot be removed by conventional Lagrange multipliers. Hence higher-order Lagrange multipliers were proposed in [2] to resolve this situation.

If a constraint is understood to be a relationship among the existing independent variables of a functional, then, since **e** is not anywhere in the principle of minimal complementary energy in (10)-(12), (4) is not a constraint of the principle of minimal complementary energy. To bring **e** into the principle of minimal complementary energy, one can artificially modify (10)-(12) using (3) to obtain the following two-field functional with the stress-strain relation as a constraint:

$$\hat{\Pi}(\boldsymbol{\sigma},\mathbf{e}) = \int_V -\frac{1}{2} a_{ijkl} e_{ij} e_{kl} dV + \int_{S_u} \bar{u}_i n_j \sigma_{ji} dS, \tag{18}$$

where

$$\sigma_{ij} = a_{ijkl} e_{kl} \quad \text{in} \quad V, \tag{19}$$

$$\sigma_{ji,j} + \bar{F}_i = 0 \quad \text{in} \quad V, \tag{20}$$

$$n_j \sigma_{ji} = \bar{p}_i \quad \text{on} \quad S_\sigma. \tag{21}$$

(18) differs from (10) by a sign which does not affect their stationary conditions. We emphasize that since (18) is a two-field functional, (19) is a relationship among existing variables and hence is a constraint. Introducing Lagrange multipliers in the usual manner, (18)-(21) lead to the following three-field functional:

$$\tilde{\Pi}(\boldsymbol{\sigma},\mathbf{e},\mathbf{u}) = \int_V [-\frac{1}{2} a_{ijkl} e_{ij} e_{kl} - u_i(\sigma_{ji,j} + \bar{F}_i) - e_{ij}(\sigma_{ij} - a_{ijkl} e_{kl})] dV$$

$$+ \int_{S_u} \bar{u}_i n_j \sigma_{ji} dS + \int_{S_\sigma} u_i(n_j \sigma_{ji} - \bar{p}_i) dS. \tag{22}$$

It is straightforward to obtain the stationary conditions of (22) as

$$\sigma_{ji,j} + \bar{F}_i = 0 \quad \text{in} \quad V, \tag{23}$$

$$\sigma_{ij} = a_{ijkl} e_{kl} \quad \text{in} \quad V, \tag{24}$$

$$e_{ij} = \frac{1}{2}(u_{i,j} + u_{j,i}) \quad \text{in} \quad V, \tag{25}$$

$$u_i = \bar{u}_i \quad \text{on} \quad S_u, \tag{26}$$

$$n_j \sigma_{ji} = \bar{p}_i \quad \text{on} \quad S_\sigma. \tag{27}$$

(22) does not have constraints. It is similar to but different from the variational functional for the generalized complementary variational principle [2] and that of the Liang-Fu principle [2].

We note that **e** may be introduced into the principle of minimal complementary energy in different ways using (3)-(6) such as

$$\frac{1}{2}b_{ijkl}\sigma_{ij}\sigma_{kl} = \frac{1}{2}a_{ijkl}e_{ij}e_{kl} = \frac{1}{2}\sigma_{ij}e_{ij} = \sigma_{ij}e_{ij} - \frac{1}{2}a_{ijkl}e_{ij}e_{kl} = \sigma_{ij}e_{ij} - \frac{1}{2}b_{ijkl}\sigma_{ij}\sigma_{kl} = \cdots \quad (28)$$

and then either (3) or (4) may be used as a constraint. Hence the above procedure of generalizing the principle of minimal complementary energy may be attempted in different ways, leading to different variational principles.

## 5. Conclusions

In the conventional form of the principle of minimal complementary energy, the strain tensor is not an independent variable and thus the stress-strain relation is not a constraint. However, the principle of minimal complementary energy may be written in a different form with the strain tensor as an independent variable and the stress-strain relation as a constraint. Then the removal of the constraint by the usual Lagrange multipliers leads to a three-field variational principle different from the existing ones in the literature. The approach in this paper, i.e., writing an existing variational functional in a different form and introducing a constraint or constraints in the process, is not unique and thus may lead to different variational principles. It is mathematical and may be used in other branches of physics.

## Acknowledgement

The result in this paper was obtained by the author in the 1980s and was mailed to H.C. Hu, the author of [5], who agreed to the above result and encouraged the author to publish it. H.C. Hu also mentioned the above result in some of his seminar(s) with recognition of the contribution of the author. The author is grateful to H.C. Hu for his encouragement to the author who was a graduate student at the time.